\documentclass[11pt]{article}
\usepackage{moriond,epsfig}
\usepackage{wrapfig}

\def\Journal#1#2#3#4{{#1} {\bf #2}, #3 (#4)}

\def\PLB{{\em Phys. Lett.}  B}

\def\EPJ{{\em Eur. Phys. J.} C}

\def\be{\begin{equation}}
\def\ee{\end{equation}}
\def\bea{\begin{eqnarray}}
\def\eea{\end{eqnarray}}

\begin{document}
\vspace*{4cm}
\title{LOW $Q^2$ STRUCTURE FUNCTIONS\\ INCLUDING THE LONGITUDINAL STRUCTURE FUNCTION}

\author{ S. SHIMIZU \\ON BEHALF OF THE H1 AND ZEUS COLLABORATIONS}

\address{Department of Physics, University of Tokyo\\
KEK IPNS, Oho 1-1, Tsukuba, Japan}

\maketitle\abstracts{
An NLO QCD analysis on the HERA combined cross sections obtained from the measurements using the data 
up to the year 2000 at both the H1 and ZEUS collaborations provides significantly improved parton 
distribution functions. In 2007, HERA successfully operated with reduced center-of-mass energies.
Direct measurements of $F_L$ were performed at $x\sim10^{-3}$,
giving a good test of the perturbative 
QCD framework in describing proton structure. 
 }

\section{Introduction}
The worlds only $ep$ collider, HERA, successfully operated from
1992 until 2007, and has played
a crucial role 
in our understanding of the proton structure. Precise measurements of the cross section of Deep 
Inelastic Scattering (DIS) were performed at HERA by both the H1 and ZEUS collaborations. 
The results revealed that the proton structure can be successfully described by 
perturbative QCD (pQCD). 
The parton distribution functions (PDFs) of the proton were well determined 
mainly based on HERA data. However, further more precise 
understanding of the proton structure is important, 
for example to describe proton-proton collisions at the LHC. In this contribution, 
the HERAPDF0.1 PDFs, 
which are the PDFs significantly improved by using
the HERA combined cross sections~\cite{heraxs},
and the direct measurements of the longitudinal structure function are presented.

\section{DIS and the proton structure}
The kinematics of lepton-proton DIS are described in terms of $x$, the Bjorken scaling variable,
$Q^2$, the virtuality of the exchanged boson,
and $y$, the fractional 
energy transfer from the lepton to the hadron system in the proton rest frame. 
Only two of them are independent at 
given centre-of-mass energy, $\sqrt{s}$, due to the relation of $Q^2=sxy$. 

The inclusive neutral current (NC) cross section of DIS, 
$ep\rightarrow e'X$, which 
proceeds via $\gamma^*/Z$ exchange,
can be written using the structure functions, $F_2$, $F_L$ and $xF_3$, as
\begin{equation}
\frac{d^2\sigma(e^\pm p)}{dxdQ^2}=\frac{2\pi\alpha^2}{Q^4}Y_+
\left[F_2(x,Q^2)-\frac{y^2}{Y_+}F_L(x,Q^2)\mp\frac{Y_-}{Y_+}xF_3(x,Q^2)\right] ,
\end{equation}
where $Y_\pm=1\pm(1-y^2)$. The structure functions reflect the momentum 
distribution of partons in the proton. They are used to determine the PDFs using the relations:
\begin{equation}
F_2=\textstyle \sum_iA_i[xq_i+x\bar q_i],\hspace{5mm}xF_3=\sum_iB_i[xq_i-x\bar q_i],
\end{equation}
where $A_i$ and $B_i$ include quark couplings to the virtual bosons. The $Q^2$ dependence of $F_2$ 
is indirectly sensitive to the gluon density. The longitudinal structure 
function, $F_L$, which is described later in this contribution, is directly sensitive to the gluon 
density but contributes to the cross section sizably only at high $y$. 

The charged current (CC) DIS, $ep\rightarrow \nu X$, proceeds via $W^\pm$ exchange 
and is a charge-selective process.
The $e^+p$ and $e^-p$ cross sections are sensitive to the 
negatively- and positively- charged quark distributions, respectively.

\section{HERA PDFs}
Both the H1 and ZEUS collaborations determined the PDFs using their own measured 
cross sections by NLO QCD analyses~\cite{h12000,zeusj}. 
Aiming for the most precise understanding of the proton structure, 
the collaborations are
working on a combination of the results to make full use of the collected data. 

The inclusive DIS cross sections were combined~\cite{heraxs} 
with the only 
assumption being
that both collaborations measure the true, hence, the same cross sections. 
The combining procedure takes full account 
of systematic correlations, leading to a significant reduction of  
systematic uncertainties. The combination was done for the low $Q^2$ 
NC cross sections from the data collected during 1996-1997 and the high $Q^2$ NC and CC 
cross sections from the 1994-2000 data.  

An NLO QCD analysis~\cite{herapdf} was performed using these HERA combined cross sections as sole input
and the PDFs were determined by a fit. The PDFs were parameterized at $Q^2_0=4\ \mathrm{GeV^2}$ 
for the valence quark, 
$xu_v$ and $xd_v$, the gluon, $xg$, and the sea quarks, 
$x\bar U=x(\bar u+\bar c)$ and $x\bar D=x(\bar d+\bar c)$, in the form of 
\begin{equation}
xf_i(x)=A_ix^{B_i}(1-x)^{C_i}(1+D_ix+E_ix^2+F_ix^3),
\end{equation}
where $D_i$, $E_i$ and $F_i$ were different from zero only if this improved the $\chi^2$ of the fit 
significantly.
In this analysis, $F_i$ was always zero and $D_i$ and $E_i$ were non zero only for $u_v$. 
Further constraints were applied and 
in total 11 free parameters describe the PDFs. They were evolved in $Q^2$ 
by the DGLAP equation at NLO 
in the $\mathrm{\overline{MS}}$ scheme with the renormalization and factorization scales taken to be 
$Q^2$. Heavy quarks were treated in the zero-mass-variable-flavour-number scheme (ZMVFN) with 
$m_c=1.4$ GeV and $m_b=4.75$ GeV.

The systematic uncertainties of the combined cross sections are small enough so that they were added 
to the statistical ones in quadrature in the fit, 
except for 4 sources arising from the combination procedure. 
The effect of these 4 sources was evaluated by the offset method~\cite{zeusj}. 

\begin{wrapfigure}[15]{r}{0.45\columnwidth}
\begin{center}
\psfig{figure=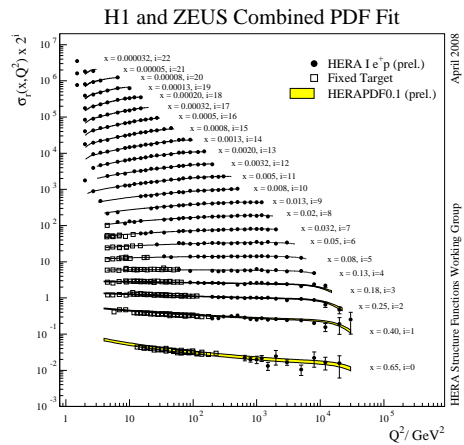,height=5.8cm}
\caption{Measured $F_2$ from HERA and fixed target experiments with the HERAPDF0.1 predictions.
\label{fig:xsec}}
\end{center}
\end{wrapfigure}
The model uncertainties of the PDFs were also evaluated. This was done by varying several 
parameters such as the heavy quark masses, the fraction of the strange and charm quarks 
among the sea quarks, the
minimum $Q^2$ cut on the data to be included and $Q^2_0$. In addition, 
further cross checks were performed. 
The value of $\alpha_s(M_Z)$ was varied for comparison.
The dependence on the choice of the parameterization was investigated by 
performing the fit 
using the H1-style~\cite{h12000} and ZEUS-style~\cite{zeusj} 
parameterizations. 

\begin{figure}
\begin{center}
\begin{tabular}{cc}
\begin{minipage}{7.9cm}
\begin{center}
\psfig{figure=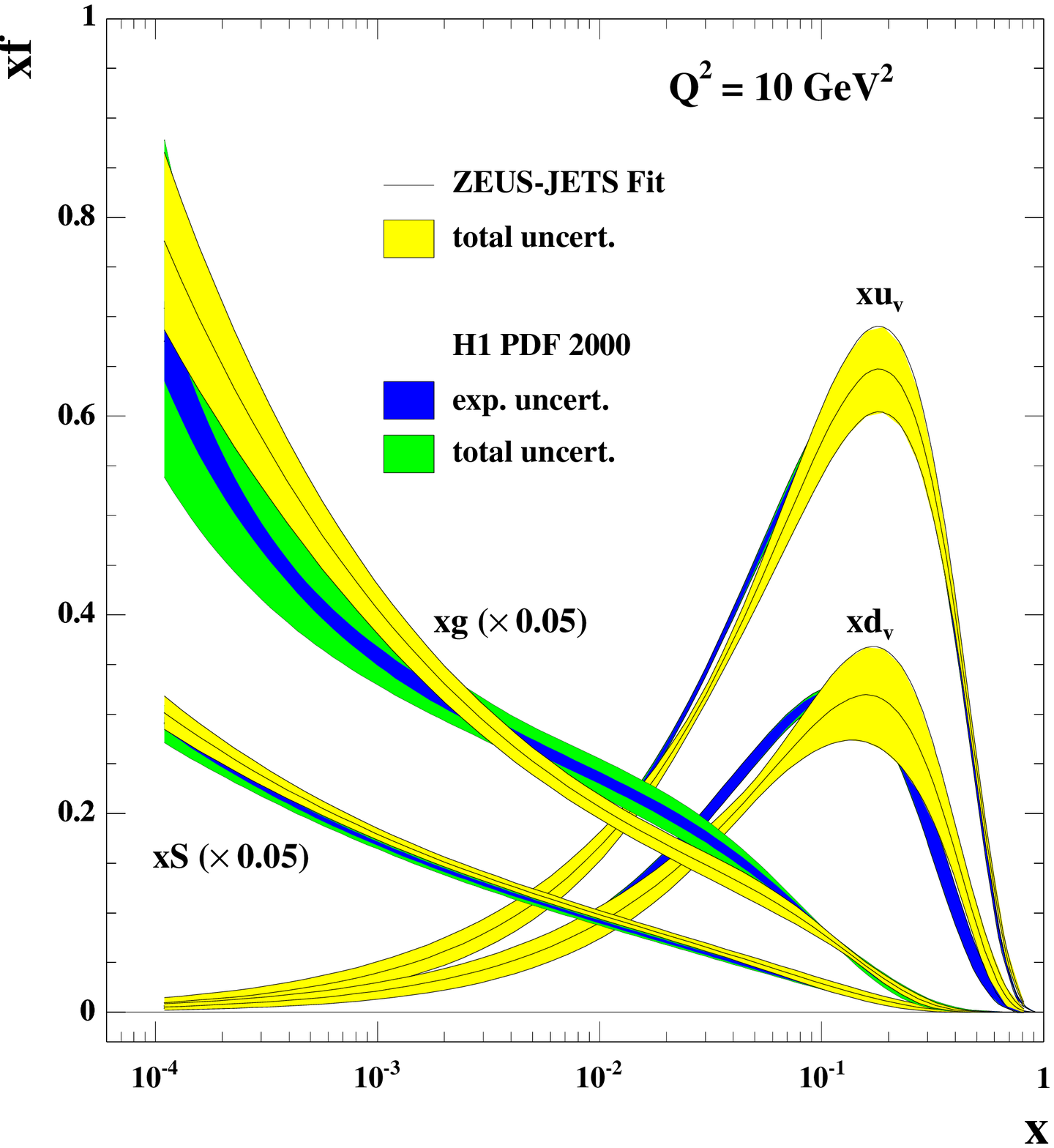,height=5.3cm}
\caption{The gluon, sea and $u$ and $d$ valence distributions
from the H1 PDF 2000 and ZEUS-JETS fits.
\label{fig:zjh1pdf}}
\end{center}
\end{minipage}
\hspace{3mm}
\begin{minipage}{7.cm}
\begin{center}
\psfig{figure=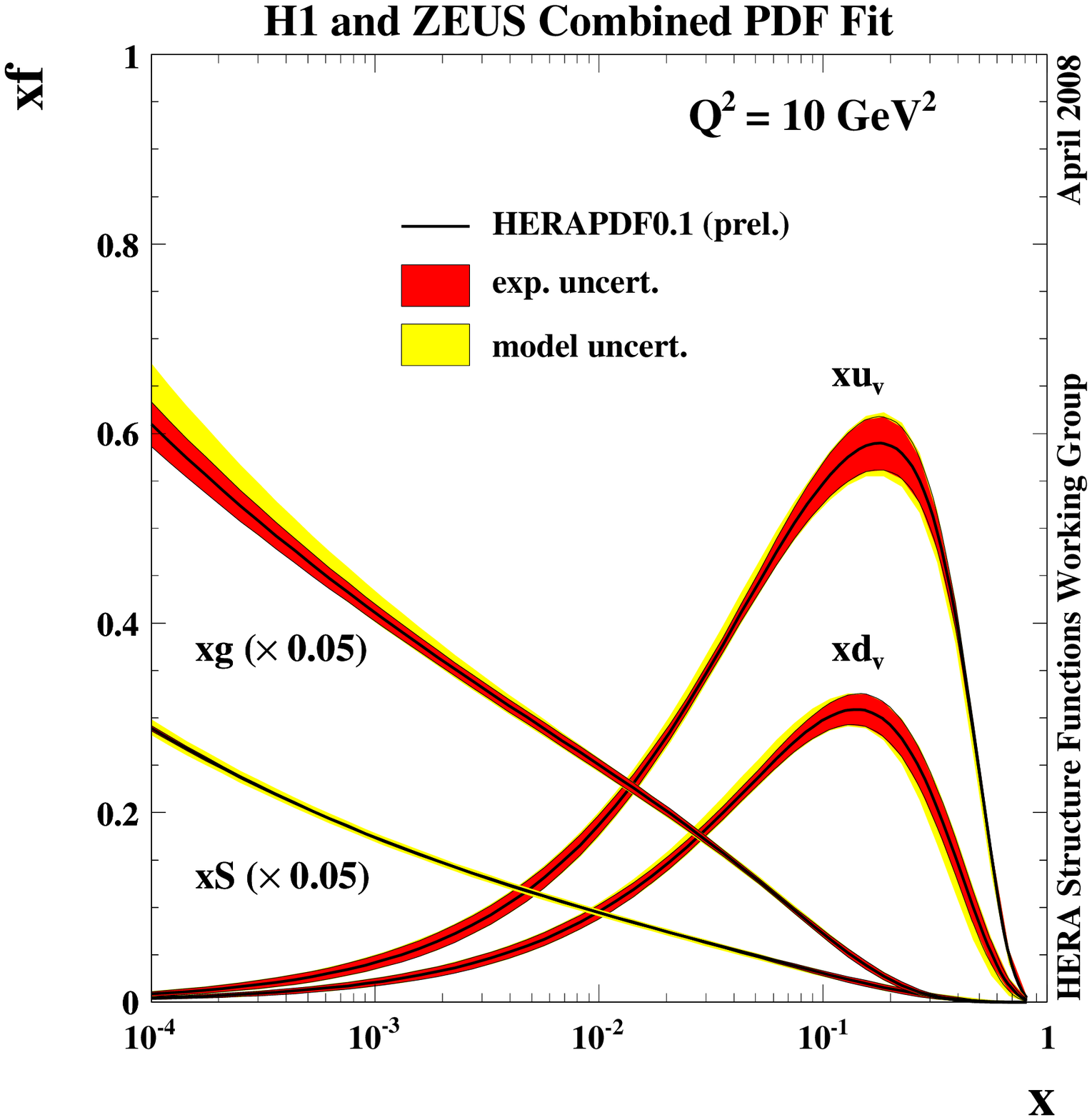,height=5.3cm}
\caption{The gluon, sea and $u$ and $d$ valence distributions from the HERAPDF0.1 fit.
\label{fig:pdf}}
\end{center}
\end{minipage}
\end{tabular}
\end{center}
\end{figure}

The resulting fit is called the HERAPDF0.1 fit. 
The data is well fitted, 
as shown in 
Fig.~\ref{fig:xsec}, which also shows that the result of the fit describes the data from 
fixed target experiments well. The PDFs 
at $Q^2=10\ \mathrm{GeV^2}$ are shown in %\nopagebreak[4]
Fig.~\ref{fig:pdf}. The uncertainties of the PDFs are impressively reduced comparing to the 
previous determinations 
using data from a single experiment. 

The H1 collaboration recently published a 
new $F_2$ result~\cite{h1f2} measured 
at $12<Q^2<150\ \mathrm{GeV^2}$ for $2\cdot10^{-4}<x<0.1\ (y<0.6)$. This result was
obtained from data collected during the years 1996-1997 and 2000. The accuracy of the measurement
is excellent ($1.5\%$ to $2\%$) and will allow a further improvement of the PDF determination. 

\section{$F_L$ measurement}
The magnitude of $F_L$ is proportional to the absorption cross section of 
the longitudinally polarized 
photons by the proton
and $F_L=0$ if the proton is composed with co-linear 
spin $\frac{1}{2}$ quarks only, as in the na\"ive QPM. However, gluon radiation in the proton 
gives non-zero values to $F_L$. Therefore, $F_L$ directly reflects 
gluon dynamics in the proton. In the PDF determination within the pQCD framework, the gluon distribution
is mainly determined from the $Q^2$ dependence of $F_2$. Since its sensitivity to the gluon 
is different, the measurement of $F_L$ is an important check for the current understanding 
of the proton structure in the pQCD framework. 

The measured variable is the reduced cross section, $\tilde\sigma$, 
which is a combination of $F_2$ and $F_L$ at low $Q^2$, 
\begin{equation}
\tilde\sigma=\frac{Q^4Y_+}{2\pi\alpha^2}\frac{d^2\sigma}{dxdQ^2}
=F_2(x,Q^2)-\frac{y^2}{Y_+}F_L(x,Q^2).
\label{eq:redxs}
\end{equation} 
The direct separation of $F_2$ and $F_L$ requires $\tilde\sigma$ measurements at the same $(x,Q^2)$
but different $y$, which means measurements at multiple beam energies. For the last three 
months of running, HERA successfully operated with significantly lowered
proton beam energies. It was the first opportunity to perform a direct $F_L$ measurement in the
low $x$ region, $x\sim10^{-3}$, where gluons are dominant in the proton. 
The measurement allows extraction of the structure functions without any QCD assumptions, 
whereas previous $F_2$ measurements use QCD calculations to estimate the $F_L$ contribution, 
as well as a consistency check of the pQCD framework for the proton structure.

Both the H1 and ZEUS collaborations collected data with $\sqrt{s}=318$, 252 and 225 GeV. 
As it can be seen from Eq.~\ref{eq:redxs}, in order to have a sizable $F_L$ contribution, 
the cross section measurement has to be done in the high $y$ 
region, where the energy of the scattered positron is small.
Mis-identification of the scattered positron introduces a large contamination of background events. 
The H1 and ZEUS collaborations took different strategies for the estimation of the contamination. The H1 
collaboration
used the event distribution of the scattered positron\nopagebreak[4]
candidates with wrong sign and ZEUS 
used Monte Calro simulation, which was validated against\nopagebreak[4]
photoproduction events identified with a dedicated electron tagger.

\begin{wrapfigure}[16]{r}{0.35\columnwidth}
\begin{center}
\psfig{figure=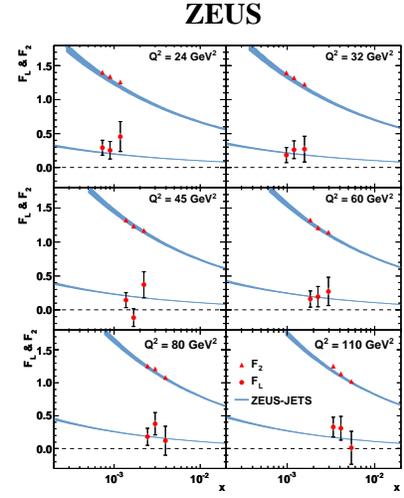,height=7.cm}
\caption{$F_2$ and $F_L$ measured by the ZEUS collaboration.
\label{fig:zeusf2fl}}
\end{center}
\end{wrapfigure}

\begin{figure}[b]
\begin{center}
\begin{tabular}{cc}
\begin{minipage}{7cm}
\psfig{figure=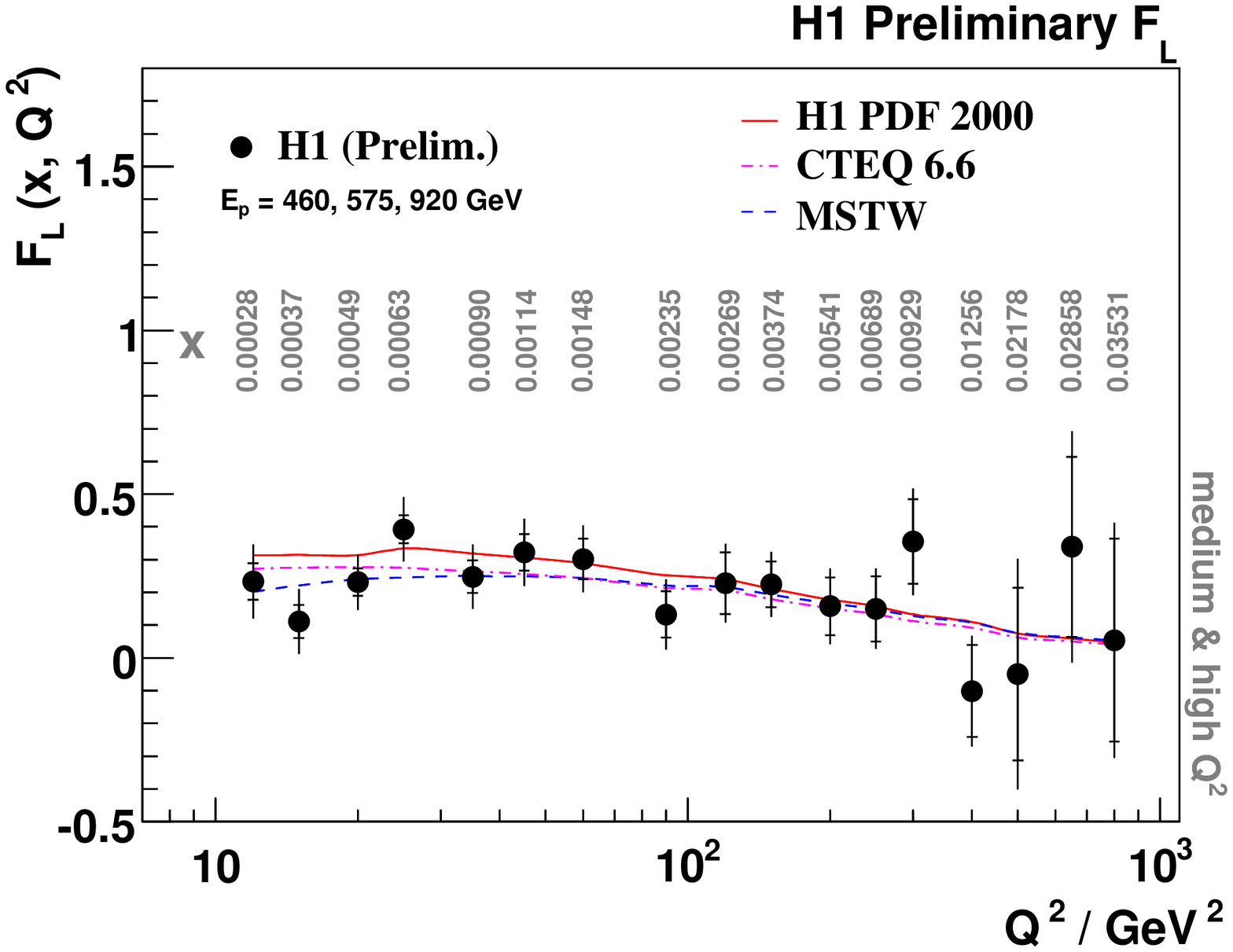,height=5cm}
\caption{Averaged $F_L$ measured by H1.
\label{fig:h1fl}}
\end{minipage}
\hspace{3mm}
\begin{minipage}{6cm}
\psfig{figure=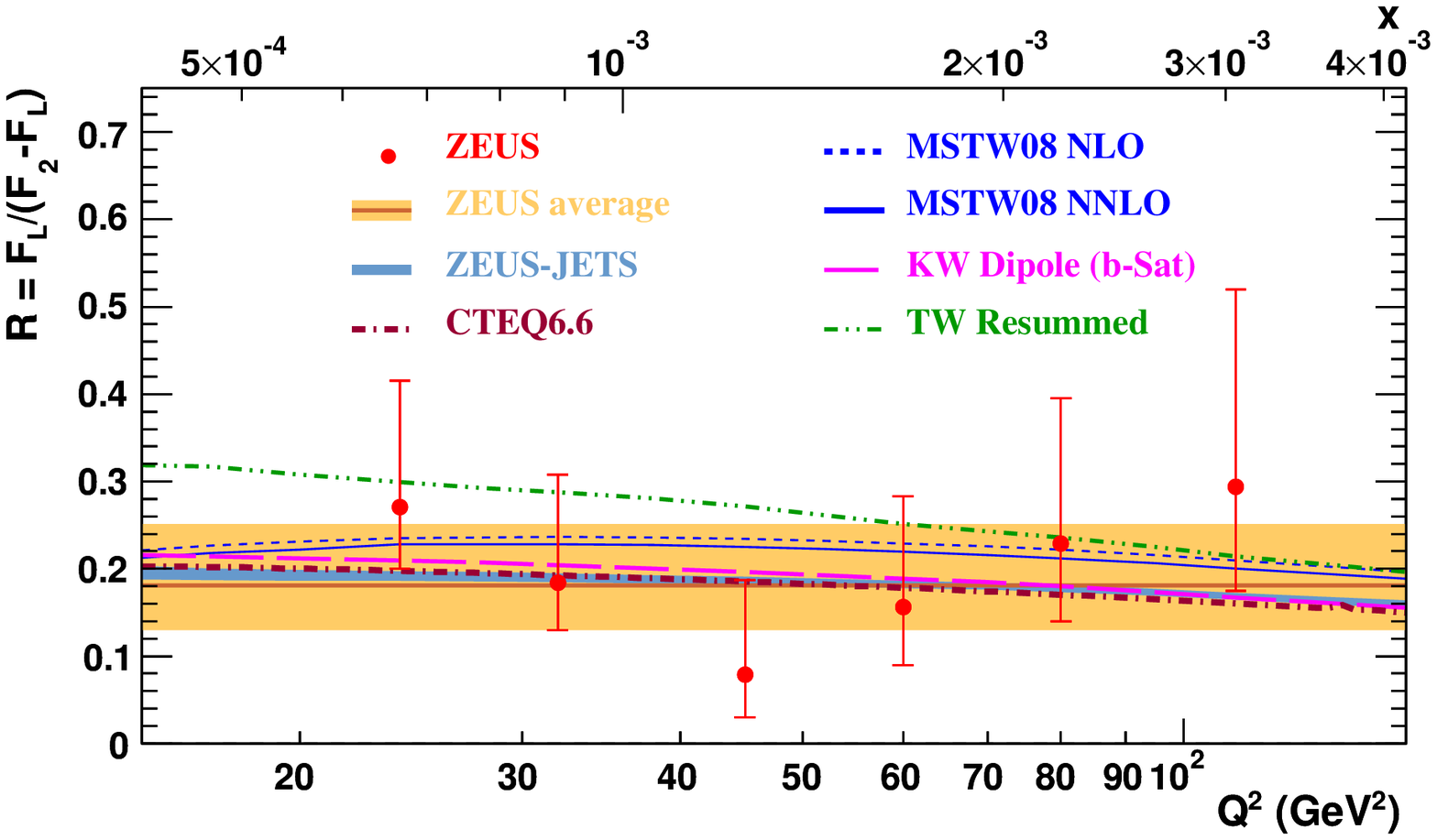,height=4cm}
\caption{$R=\frac{F_L}{F_2-F_L}$ measured by ZEUS.
\label{fig:zeusr}}
\end{minipage}
\end{tabular}
\end{center}
\end{figure}

The DIS reduced cross sections were measured at each beam energy for 
$12<Q^2<90\ \mathrm{GeV^2}$ and 
$35<Q^2<800\ \mathrm{GeV^2}$ by H1~\cite{h1fl,h1flhq} and 
$24<Q^2<110\ \mathrm{GeV^2}$ by ZEUS~\cite{zeusfl}. 
The extraction of $F_L$ was performed using these reduced cross section measurements.  
Figure~\ref{fig:zeusf2fl} shows directly separated $F_2$ and $F_L$ by ZEUS. 
A non-zero $F_L$, $0<F_L<F_2$, is supported by the measurement. In the figure, the measurements 
are compared to the pQCD prediction based on the ZEUS-JETS PDFs~\cite{zeusj}. The prediction is 
consistent with the measurement. 
Figures~\ref{fig:h1fl}~and~\ref{fig:zeusr} show $F_L$ and $R=\frac{F_L}{F_2-F_L}$ averaged 
in each $Q^2$ bin at a given $x$ from H1 and ZEUS, respectively. 
The measurements are compared to predictions using several PDFs and pQCD frameworks 
but are not sensitive 
to the differences between them. The measurements are consistent with all the predictions. 
An overall value of $R$ was provided by ZEUS as $R=0.18^{+0.07}_{-0.05}$.

\section{Summary}
Electron-proton scattering at HERA provides precise information on the structure of the proton. 
The HERAPDF0.1 PDFs, obtained using the combined cross sections of the H1 and ZEUS collaborations, 
have impressive precision, much improved with respect to the PDFs determined by the individual collaborations.
The direct measurements of $F_L$ were performed.
For the first time, $F_L$ was separated from $F_2$ without any QCD assumption. A non-zero 
$F_L$ is supported. The measurements are in good agreement with existing\nolinebreak[4] models.

%\section*{Acknowledgments}


\section*{References}
\begin{thebibliography}{99}
%\bibitem{ja}C Jarlskog in {\em CP Violation}, ed. C Jarlskog
%(World Scientific, Singapore, 1988).
%\bibitem{ma}L. Maiani, \Journal{\PLB}{62}{183}{1976}.
%\bibitem{bu}J.D. Bjorken and I. Dunietz, \Journal{\PRD}{36}{2109}{1987}.
%\bibitem{bd}C.D. Buchanan {\it et al}, \Journal{\PRD}{45}{4088}{1992}.

\bibitem{heraxs} H1 and ZEUS Coll., H1prelim-07-007, ZEUS-prel-07-026. 
\bibitem{h12000}H1 Coll., C. Adloff {\it et al.}, \Journal{\EPJ}{30}{1}{2003}.
\bibitem{zeusj}ZEUS Coll., S. Chekanov {\it et al.}, \Journal{\EPJ}{42}{1}{2005}.
\bibitem{herapdf}A.M. Cooper-Sarker, Proc. of DIS2008, London, 
\verb$doi:10.3360/dis.2008.25$
%XVI Int. Workshop on Deep-Inelastic 
%Scattering and Related Topics, London, England, April 2008, 
%\verb$http://dx.doi.org/10.3360/dis.2008.25$
\bibitem{h1f2}H1 Coll., F.D. Aaron {\it et al.}, DESY-09-005, 
to be published in {\it Eur. Phys. J.} C.
\bibitem{h1fl}H1 Coll., F.D. Aaron {\it et al.}, \Journal{\PLB}{665}{139}{2008}.
\bibitem{h1flhq}V. Chekelian, Proc. of DIS2008, London, 
\verb$doi:10.3360/dis.2008.39$
%XVI Int. Workshop on Deep-Inelastic 
%Scattering and Related Topics, London, England, April 2008, 
%\verb$http://dx.doi.org/10.3360/dis.2008.39$
\bibitem{zeusfl}ZEUS Coll., S. Chekanov {\it et al.}, DESY-09-046, 
to be published in {\it Phys. Lett.} B.


\end{thebibliography}
\end{document}